\documentclass[showpacs,apl,twocolumn,floatfix]{revtex4}

\usepackage{graphicx,float,amsmath}
\usepackage{color}

\newcommand{\specialcell}[2][c]{%
  \begin{tabular}[#1]{@{}c@{}}#2\end{tabular}}

\begin{document}

\title{Polarizers, optical bridges and Sagnac interferometers for nanoradian polarization rotation measurements}
\author{A.C.H. Rowe}
\thanks{alistair.rowe@polytechnique.edu}
\author{I. Zhaksylykova}
\author{G. Dilasser}
\author{Y. Lassailly}
\author{J. Peretti}
\affiliation{Laboratoire de Physique de la Mati\`ere Condens\'ee, Ecole Polytechnique, CNRS, Universit\'e Paris Saclay, 91128 Palaiseau Cedex, France}

\begin{abstract}
The ability to measure nanoradian polarization rotations, $\theta_F$, in the photon shot noise limit is investigated for partially crossed polarizers (PCP), a static Sagnac interferometer and an optical bridge, each of which can in principal be used in this limit with near equivalent figures-of-merit (FOM). In practice a bridge to PCP/Sagnac source noise rejection ratio of $1/4\theta_F^2$ enables the bridge to operate in the photon shot noise limit even at high light intensities. The superior performance of the bridge is illustrated via the measurement of a 3 nrad rotation arising from an axial magnetic field of 0.9 nT applied to a terbium gallium garnet. While the Sagnac is functionally equivalent to the PCP in terms of the FOM, unlike the PCP it is able to discriminate between rotations with different time ($T$) and parity ($P$) symmetries. The Sagnac geometry implemented here is similar to that used elsewhere to detect non-reciprocal ($\overline{T}P$) rotations like those due to the Faraday effect. Using a Jones matrix approach, novel Sagnac geometries uniquely sensitive to non-reciprocal $\overline{TP}$ (e.g. magneto-electric or magneto-chiral) rotations, as well as to reciprocal rotations (e.g. due to linear birefringence, $TP$, or to chirality, $T\overline{P}$) are proposed.
\end{abstract}
\maketitle

\section{Introduction}

The relative phase and intensity of the two component polarizations of light may be modified during transmission according to the spectral details and symmetries of the transmission medium's electronic structure, and thus the ability to sensitively measure the corresponding rotations and/or ellipticities can be used to probe this structure. Polarization rotations may arise through a number of physically distinct phenomena that are characterized by their time ($T$) and parity ($P$) symmetries, of which there are four possible combinations. Time anti-symmetric, inversion symmetric ($\overline{T}P$) rotations include the Faraday effect in magnetized media, with examples of small angle rotation measurements including the study of magnetic anisotropy\cite{su2013}, single spin dynamics in quantum wells \cite{atature2007}, magnetic vortices in superconductors \cite{koblischka1995}, non-equilibrium spin polarized electrons in semiconductors \cite{kikkawa99,kato2004} and orbital magnetism in graphene \cite{crassee2011}. Measurement of $T\overline{P}$ rotations include those in chiral media such as liquids and thin films \cite{chou1997}, as well as in chiral metamaterials \cite{plum2009}. $TP$ rotations include those due to linear birefringence, for example arising in optically transparent biological tissue \cite{takeguchi1968}, while $\overline{TP}$ rotations include those arising from a range of novel phenomena or materials including ferroelectricity \cite{krichevtsov1996}, proposed pseudo-gap states in cuprate superconductors \cite{xia2006b} and in chiral spin liquids \cite{kleindienst1998, vallet2001, machida2010}.

Polarization rotation measurements have been achieved using a number of optical configurations including the standard partially crossed polarizers (PCP), a variety of optical bridges \cite{chang2011, li2014}, and most recently modified Sagnac interferometers \cite{xia2006}. Here we compare and contrast these techniques in terms of their ultimate theoretical and practical sensitivities to nanoradian polarization rotations, and also in terms of their ability to distinguish rotations arising in media which break or preserve time and/or parity symmetries. In order to compare their respective sensitivities, simple expressions are given for the root-mean-square (RMS) noise, the signal-to-noise ratio (SNR) and a figure-of-merit (FOM) defined as the squared SNR per unit bandwidth. The validity of these expressions is demonstrated experimentally by performing Faraday rotation ($\overline{T}P$) measurements in the PCP, bridge and Sagnac geometries on a terbium gallium garnet (TGG) rod exposed to a small, time varying magnetic field. By changing the amplitude of the oscillating magnetic field, Faraday rotations from 38 $\mu$rad down to 3 nrad are explored, and it will be shown that thanks to a large bridge to PCP source noise rejection ratio, the bridge is the preferred method to achieve nanoradian rotation measurements in the shot noise limit. Although rotations arising from $T\overline{P}$, $TP$ and $\overline{TP}$ phenomena are not experimentally considered, the results obtained here concerning the FOM are applicable to these cases.

Although the Sagnac interferometer will be shown to have a lower FOM than either PCP or the optical bridge, it is capable (in a single measurement) of discriminating between rotations arising from different phenomena with different time and/or spatial symmetries \cite{leilabady1986}. Sagnacs have been used to discriminate the magneto-optical effect ($\overline{T}P$) from rotations due to linear birefringence ($TP$) \cite{spielman1990} and since then the use of Sagnac interferometers in magneto-optics \cite{kapitulnik1994, beyersdorf1999, xia2006} has become synonymous with the effort to experimentally discriminate between reciprocal ($T$) from non-reciprocal ($\overline{T}$) phenomena that result in polarization rotations. Using the constraints imposed on the form of the Jones' matrix by the time and spatial symmetries of the optical medium \cite{armitage2014}, it is shown here that a range of modified Sagnac interferometers exist which can be configured to be sensitive to rotations arising from only one of the $TP$, $\overline{T}P$, $T\overline{P}$ or $\overline{TP}$ symmetric media. Schematic diagrams of four such novel Sagnac loops are given, and the functional details of each are discussed.   

\section{Signal-to-noise ratio in the measurement of Faraday effect rotations}
\label{Theory}

In order to demonstrate the capabilities of PCP, Sagnacs and bridges when measuring nano-rotations, the Faraday effect in a TGG will be used. The Faraday effect results from the dependence of the complex dielectric constant on the sample magnetization \cite{halpern1964,freiser1968} and a simplified description of its effect on the transmitted light polarization and ellipticity is obtained using the Jones' matrix formalism \cite{goldstein2010}. The general form of the Jones' matrix describing the polar Faraday effect is \cite{armitage2014}: \begin{equation} \label{FaradayJones} \textbf{F}(A,B,D) = \left[ \begin{array}{cc} A & -B \\ B & D \end{array} \right], \end{equation} where $A, B, D$ may be complex. The inequality $A \neq D$ describes the Faraday-induced ellipticity so that in the case of a pure, right-handed Faraday rotation through an angle $\theta_F$, $A = D = \cos\theta_F$, $B = \sin\theta_F$ and $\textbf{F}(A,B,D)$ may be written $\textbf{F}(\theta_{F})$. Here the boldface is used to denote a matrix operator and it is noted that all Jones' matrices will be written in the reference frame of the light i.e. in which the z-direction is parallel to the photon wavevector. In the Jones' matrix description of the Sagnac interferometers this will have consequences for the forms of the matrices describing each optical element when traversed by either the clockwise (CW) or the counter clockwise (CCW) propagating beams \cite{armitage2014}. Moreover in the following, reflections at interfaces and light absorption will be neglected for the sake of simplicity. 

\subsection{Ideal description of partially crossed polarizers (PCP)}
\label{PCPTheory}

\begin{figure}[tbp]
\includegraphics[clip,width=8.5 cm] {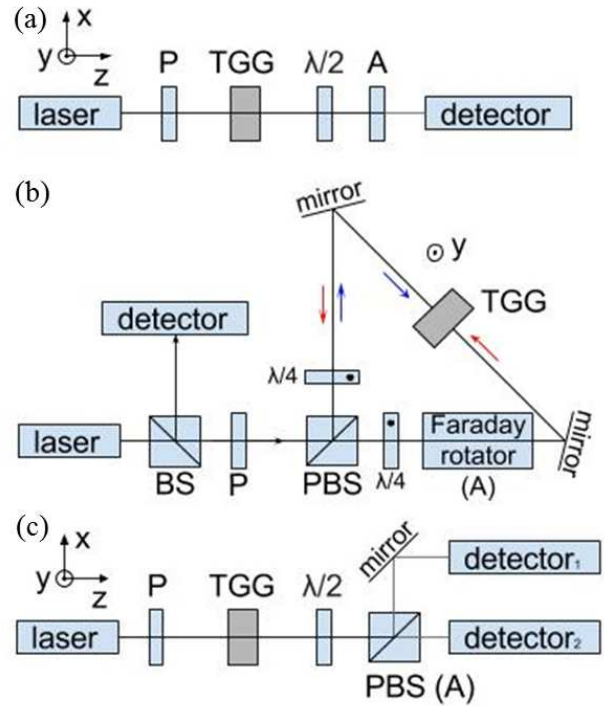}
\caption{Schematics of (a) the PCP experiment, (b) the Sagnac interferometer, and (c) the optical bridge. In these schematics diagrams, source polarizers are labeled ``P'', linear analyzers ``A'', non-polarizing beam splitters ``BS'', polarizing beam splitters ``PBS'', and half and quarter wave plates $\lambda/2$ and $\lambda/4$ respectively. The terbium gallium garnet used to create a small Faraday rotation is labeled ``TGG''.}
\label{configs}
\end{figure}

In the simplest geometry shown in Fig. \ref{configs}(a) a linearly polarized source of intensity $I_0$, polarized in the y-direction by the polarizer labeled P, is transmitted through the sample before being analyzed by the combination of the $\lambda$/2 waveplate and a linear analyzer (A). Here intensities are photon fluxes measured in photons/s normalized to a unit area. In terms of the Jones' vectors and matrices $I_0 \propto |\vec{E}|^2 = |(0,E_y)|^2$ where $E_y$ is the y-component of the incident electric field (the propagation direction is taken, by definition, to be the +z direction), while the $\lambda$/2 waveplate is described by $\textbf{PR}(\pi,0,\theta/2)$ and the analyzer (aligned in the x-direction i.e. crossed with the polarizer, P) is described by $\textbf{P}(0)$. The definitions of the Jones' matrix for a generalized phase retarder $\textbf{PR}(\phi_x,\phi_y,q)$ and a polarizer $\textbf{P}(\theta_P)$ are given in Appendix \ref{Jones}. Here $\theta$ is the angle that the fast axis of the $\lambda$/2 plate makes with the x-direction and it will turn out to be the effective analyzer angle for the linear polarization analysis.  In the case of a sample resulting in a Faraday rotation of $-\theta_F$, the Jones' vector at the detector is then $\textbf{P}(\pi/2).\textbf{PR}(\pi,0,\theta/2).\textbf{F}(-\theta_F).\vec{E}$, corresponding to an intensity on the detector given by the modulus squared of this vector: \begin{equation} \label{Malus} I_{\textrm{det}} = I_0\sin^2(\theta + \theta_F). \end{equation} This is just a modified version of Malus' law where, in the absence of a rotation $-\theta_F$, $\theta = 0$ corresponds to the crossed polarizers position.

In a standard Faraday rotation measurement, the sign of the polarization rotation is periodically reversed by reversing the component of the sample magnetization in the direction of the light propagation \cite{freiser1968}. In this reversed measurement, the intensity received on the detector is $I_0\sin^2(\theta - \theta_F)$. The  magneto-optical (MO) signal can then be defined as the difference obtained between these two measurements: \begin{align} \label{PCPcontrast} \Delta I & = I_0\left\{\sin^2(\theta + \theta_F)-\sin^2(\theta - \theta_F)\right\} \notag \\ & = 2I_0\sin\theta_F\sin2\theta. \end{align} 

The noise on the detector during these measurements can be classified according to its origin: i) electronic noise whose RMS intensity is independent of the light intensity on the detector. Experiments should be designed so that this component is negligible, and it will not be mentioned further here, ii) the photon source noise arising from quantum noise associated with the spontaneous emission in the gain medium, mechanical vibrations of optical components in the beam path etc..., whose RMS is proportional, via a factor $\mathcal{B}$ measured in $1/\sqrt{\textrm{Hz}}$, to the light intensity on the detector, i.e. \begin{equation} \label{sourcenoise} \textrm{N}_{\textrm{so}}=\mathcal{B}I_{\textrm{det}}\sqrt{\Delta f}. \end{equation} Here $\Delta f$ is the detection bandwidth. The physical significance of $\mathcal{B}$ will be discussed in Section \ref{Experiment}, and iii) the intrinsic photon shot noise related to the Poissonian distribution of the arrival times of individual photons at the detector, whose RMS varies as the square root of the intensity on the detector i.e. \begin{equation} \label{shotnoise} \textrm{N}_{\textrm{sh}}=\sqrt{I_{\textrm{det}}\Delta f}. \end{equation} In the absence of specialized techniques related to quantum squeezing \cite{teich1989}, one usually wishes to work in the photon shot noise limit since this is the intrinsic noise floor in a standard optical experiment. At the same time one should maximize $I_0$ since this maximizes MO signal, Eq. \ref{PCPcontrast}. 

Using the intensities calculated above for the two rotations $\pm\theta_F$ along with the expressions in Eq. \ref{sourcenoise} and Eq. \ref{shotnoise}, it is possible to calculate expressions for the RMS photon source and shot noise on the MO signal using the error propagation formula (according to which the total noise variance is the sum of the variances corresponding to each uncorrelated noise source). In this case the RMS source noise on the MO signal measurement is: \begin{equation} \label{PCPsource} \textrm{N}_{\textrm{PCP,so}} = \mathcal{B}I_0\sqrt{\Delta f}\sqrt{\sin^4(\theta+\theta_F)+\sin^4(\theta-\theta_F)}, \end{equation} while the RMS shot noise on the MO signal is: \begin{equation} \label{PCPshot} \textrm{N}_{\textrm{PCP,sh}} = \sqrt{2I_0\Delta f(\sin^2\theta\cos^2\theta_F+\cos^2\theta\sin^2\theta_F)}. \end{equation} 

The source and shot noise limited SNR is then given by the ratio of Eq. \ref{PCPcontrast} to Eq. \ref{PCPsource} and Eq. \ref{PCPshot} respectively, yielding: \begin{equation} \label{PCPsourceSNR} \textrm{SNR}_{\textrm{PCP,so}} = \frac{2\sin\theta_F\sin2\theta}{\mathcal{B}\sqrt{\Delta f}\sqrt{\sin^4(\theta+\theta_F)+\sin^4(\theta-\theta_F)}} \end{equation} and \begin{equation} \label{PCPshotSNR} \textrm{SNR}_{\textrm{PCP,sh}} = \frac{\sqrt{2I_0}\sin\theta_F\sin2\theta}{\sqrt{\Delta f}\sqrt{\sin^2\theta\cos^2\theta_F+\cos^2\theta\sin^2\theta_F}}. \end{equation} Note that $\textrm{SNR}_{\textrm{PCP,so}}$ is independent of $I_0$ while $\textrm{SNR}_{\textrm{PCP,sh}}$ varies as $\sqrt{I_0}$, meaning that at sufficiently low incident intensities the limiting optical noise source (i.e. that yielding the lowest SNR) will always be the photon shot noise. However, reducing $I_0$ reduces the absolute value of the shot noise SNR so obtaining a shot noise limited measurement in this way is not generally recommended.

\subsection{Ideal description of the static Sagnac interferometer}
\label{SagnacTheory}

Figure \ref{configs}(b) shows the modified Sagnac interferometer that will be considered (and implemented) here. It is a static version of the loop employed elsewhere \cite{xia2006} in which the static Faraday rotator replaces an electro-optic modulator. While this has consequences for the sensitivity of the loop to rotations other than Faraday ($\overline{T}P$) rotations (see Section \ref{SagnacGeometries}), the principal role of these elements is to impose a particular analysis angle, $\theta$. This becomes apparent when the Jones' matrix calculation is carried out for the clockwise (CW) and counter-clockwise (CCW) beams circulating in the Sagnac loop. 

Consider first the CW beam. The Jones' vector due to the CW beam (blue arrow in Fig. \ref{configs}(b)) appearing at the detector is given by \begin{align} \label{CW} \vec{E}_{\textrm{CW}} = & \textbf{ABSR}.\textbf{Y}^{-1}_\pi.\textbf{P}^{\textrm{T}} (\tfrac{\pi}{4}).\textbf{Y}_\pi.\textbf{Y}^{-1}_\pi.\textbf{P}^{\textrm{T}}(0).\textbf{Y}_\pi. \notag \\ & \textbf{PR}(\tfrac{\pi}{2},0,\tfrac{\pi}{4}).\textbf{F}(\tfrac{\pi}{2}-\theta).\textbf{F}(-\theta_F).\textbf{PR}(\tfrac{\pi}{2},0,\tfrac{\pi}{4}). \notag \\ & \textbf{P}(\tfrac{\pi}{2}).\textbf{P}(\tfrac{\pi}{4}).\textbf{BST}.\vec{E} \end{align} where $\vec{E} = (0,E_y)$ as before. This rather inelegant expression can be broken down into its individual parts to better understand the role of each component in Fig. \ref{configs}(b). The non-polarizing beam splitter (BS) is represented by four separate Jones' matrices corresponding to transmission from an input port to an output port ($\textbf{BST}$), reflection from an input port to an output port ($\textbf{BSR}$), reverse transmission (i.e. from an output port to an input port, $\textbf{ABST}$), and reverse reflection (i.e. from an output port to an input port, $\textbf{ABSR}$). These matrices were determined experimentally and are given in Appendix \ref{Jones}. In this interferometer the BS is outside the Sagnac loop and serves only to redirect the interfering light to the detector. The second component outside the Sagnac loop is a linear polarizer aligned at 45 degrees to the x-axis, $\textbf{P}(\pi/4)$. With respect to the axes of the polarizing beam splitter (PBS) that forms the Sagnac loop, this polarizer results in equal powers circulating in the CW and CCW directions. This is because the PBS reflects (transmits) only the y-component (x-component) of the incident face. In the case of the reflection this is represented by $\textbf{P}(\pi/2)$ while transmission through the PBS is given by $\textbf{P}(0)$. Two quarter wave plates are placed in the loop with their fast axes aligned at 45 degrees to the x-axis, $\textbf{PR}(\pi/2,0,\pi/4)$, so that in the Sagnac loop between these waveplates the light circulates in a circularly polarized state. The variable (but static) Faraday rotator rotates the CW light polarization through an angle $\pi/2 - \theta$ according to $\textbf{F}(\pi/2-\theta)$ where $\theta$ will turn out to be the same analysis angle used in the description of the PCP. The sample itself causes a rotation $-\theta_F$ described by $\textbf{F}(-\theta_F)$.

There are a number of subtle points to note. Firstly, the mirrors shown in Fig.\ref{configs}(b) were ignored in Eq. \ref{CW}. This can be done in a free-space (i.e. mirrored) Sagnac for an \textit{even} number of mirrors as long as the light is approximately normally incident on each mirror i.e. far from the Brewster angle. The Jones' matrix, $\textbf{M}$, for an ideal mirror and an incident angle far from the Brewster angle, is given in Appendix \ref{Jones}, from which it can be seen that $\textbf{M}.\textbf{M}=\textbf{1}$ where $\textbf{1}$ is the $2\times 2$ identity matrix. Eq. \ref{CW} is strictly correct for a fibered loop which contains no mirrors. Secondly, when the light passes through optical components via their back face, the Jones' matrix needs to be modified by transposing it \cite{armitage2014}. Furthermore since the coordinate axis system is defined in the reference frame of the light and not the laboratory, a rotation of the Jones' matrix about the y-axis by $\pi$ radians is required (given by the operator, $\textbf{Y}_\pi$). If the optical component in question is reciprocal then no further operation is required, whereas for non-reciprocal components the sign of the off diagonal elements of the Jones' matrix must be reversed \cite{armitage2014}. Thus the term $\textbf{Y}^{-1}_\pi.\textbf{P}^{\textrm{T}} (\tfrac{\pi}{4}).\textbf{Y}_\pi$ in Eq. \ref{CW}, where the superscript $\textrm{T}$ refers to the transpose operation, corresponds to the retransmission of the light through the polarizer, P, after exiting the Sagnac loop. In the same way $\textbf{Y}^{-1}_\pi.\textbf{P}^{\textrm{T}}(0).\textbf{Y}_\pi$ corresponds to the backwards transmission of the CW light through the PBS.

The same treatment, included here for completeness, can be given for the CCW propagating beam (red arrow in Fig. \ref{configs}(b)): \begin{align} \label{CCW} \vec{E}_{\textrm{CCW}} = & \textbf{ABSR}.\textbf{Y}^{-1}_\pi.\textbf{P}^{\textrm{T}} (\tfrac{\pi}{4}).\textbf{Y}_\pi.\textbf{Y}^{-1}_\pi.\textbf{P}^{\textrm{T}}(\tfrac{\pi}{2}).\textbf{Y}_\pi. \notag \\ & \textbf{PR}(\tfrac{\pi}{2},0,\tfrac{-\pi}{4}).\textbf{F}(\theta_F).\textbf{F}(\tfrac{-\pi}{2}+\theta).\textbf{PR}(\tfrac{\pi}{2},0,\tfrac{-\pi}{4}). \notag \\ & \textbf{P}(0).\textbf{P}(\tfrac{\pi}{4}).\textbf{BST}.\vec{E} \end{align} where the changes to the reciprocal waveplates and the non-reciprocal Faraday rotations are the result of the operations required for a reversal of the light propagation outlined immediately above and described in detail elsewhere \cite{armitage2014}. Notice in particular that the Faraday rotation due to the sample is described by $\textbf{F}(\theta_F)$ i.e. a rotation in the opposite direction to that experienced by the CW beam. At first sight this seems surprising since the sign of the Faraday rotation depends only on the direction of the sample magnetization (which has not changed). However, the sample magnetization does not change in the laboratory reference frame, whereas it \textit{has} changed sign in the reference frame of the light (i.e. for CW and CCW beams).

The resultant Jones' vector describing the interference on the detector is then found by summing Eq. \ref{CW} and Eq. \ref{CCW}. The modulus squared of this vector gives the detected light intensity: \begin{equation} \label{SagnacMalus} I_{\textrm{det}} = \frac{I_0}{8}\sin^2(\theta + \theta_F), \end{equation} whose angular dependence is identical to Malus' result found for the PCP (Eq. \ref{Malus}). The reduction in detected intensity by a factor of 8 relative to PCP results from the loss of photons at the beam splitters. The MO signal on the Sagnac is therefore also a factor of 8 smaller than for PCP: \begin{equation} \label{Sagnaccontrast} \Delta I = \frac{I_0}{4}\sin\theta_F\sin 2\theta, \end{equation} although its functional dependence on $\theta$ and $\theta_F$ are identical. Since the RMS source noise in Eq. \ref{PCPsource} is linear in $I_0$, the source noise in the Sagnac is also reduced by a factor of 8 and the source noise SNR is identical to that of the PCP (see Eq. \ref{PCPsourceSNR}). On the other hand, the RMS shot noise, Eq. \ref{PCPshot} varies as $\sqrt{I_0}$ so that it becomes \begin{equation} \label{Sagnacshot} \textrm{N}_{\textrm{S,sh}} = \frac{\sqrt{I_0\Delta f(\sin^2\theta\cos^2\theta_F+\cos^2\theta\sin^2\theta_F)}}{2}\end{equation} such that the shot noise limited SNR in the Sagnac is \begin{equation} \label{SagnacshotSNR} \textrm{SNR}_{\textrm{S,sh}} = \frac{\sqrt{I_0}\sin\theta_F\sin2\theta}{2\sqrt{\Delta f}\sqrt{\sin^2\theta\cos^2\theta_F+\cos^2\theta\sin^2\theta_F}},\end{equation} a factor of $\sqrt{8}$ smaller than the PCP. For a given laser source intensity, $I_0$, the Sagnac SNR will therefore be smaller than the PCP in the shot noise limit. If the measurement is source noise limited, the SNR for the PCP and Sagnac will, in principal, be identical. As will be discussed in Section \ref{SagnacGeometries}, Sagnac interferometers also differ from PCP in that their symmetry can be adapted to polarization rotations arising from optical phenomena with specific time and spatial symmetries.

\subsection{Ideal description of the optical bridge}
\label{BridgeTheory}

The linear analyzer in the PCP is replaced by a PBS in the optical bridge as shown in Fig. \ref{configs}(c). As such, both the x- and y-components of the light transmitted through the sample are used. With the two photodiode detectors connected back-to-back, the resulting output signal is proportional to their difference intensity. For a Faraday rotation $-\theta_F$, the Jones' vector of the x-channel light is $\textbf{P}(0).\textbf{PR}(\pi,0,\theta/2).\textbf{F}(-\theta_F).\vec{E}$, while that on the y-channel is $\textbf{P}(\pi/2).\textbf{PR}(\pi,0,\theta/2).\textbf{F}(-\theta_F).\vec{E}$. The difference of the squared modulus of each of these vectors gives the output of the optical bridge: \begin{equation} \label{Bridgeoutput} I_{\textrm{bridge}} = I_0\cos2(\theta+\theta_F). \end{equation} As in the case of the PCP and the Sagnac, the MO signal is then obtained by reversing the magnetization of the sample to obtain a Faraday rotation of $\theta_F$, yielding a bridge output of $I_0\cos2(\theta-\theta_F)$. The difference in the responses for Faraday rotations of $\pm\theta_F$ is the MO signal: \begin{equation} \label{Bridgecontrast} \Delta I = 2I_0\sin2\theta\sin2\theta_F. \end{equation}

Since the optical source noise on each of the detectors is correlated, it is (partially) removed during the difference measurement on the optical bridge. It is therefore the difference intensity on the bridge, given by Eq. \ref{Bridgeoutput} for a rotation of $+\theta_F$, which is relevant for the source noise calculation rather than the individual intensities on each detector. Using the error propagation formula for the two rotations, $\pm\theta_F$, the RMS source noise on the MO signal is then \begin{align} \label{Bridgesource} \textrm{N}_{\textrm{B,so}} &= \mathcal{B}I_0\sqrt{\Delta f}\sqrt{\cos^22(\theta+\theta_F)+\cos^22(\theta-\theta_F)} \notag \\ &= \mathcal{B}I_0\sqrt{\Delta f}\sqrt{1+\cos4\theta\cos4\theta_F}. \end{align} It is interesting to note that for $\theta = \pi/4$, $\textrm{N}_{\textrm{B,so}}$ is minimized and reduces to $\mathcal{B}I_0\sqrt{\Delta f}$ while the MO signal, Eq. \ref{Bridgecontrast}, is maximized. The shot noise on the other hand, being uncorrelated on each of the two detectors, must be calculated differently. For a given rotation angle, for example $+\theta_F$, the total shot noise is found by using the error propagation formula with the intensities on each of the two detectors which yields $\sqrt{I_0(\sin^2(\theta-\theta_F)+\cos^2(\theta-\theta_F))\Delta f} = \sqrt{I_0\Delta f}$. The same result is obtained when the rotation angle is $-\theta_F$ so that the shot noise on the MO signal is \begin{equation} \label{Bridgeshot} \textrm{N}_{\textrm{B,sh}} = \sqrt{2I_0\Delta f}. \end{equation}

The SNR for each of the two cases, source and shot, is then calculated by taking the ratio of Eq. \ref{Bridgecontrast} to Eq. \ref{Bridgesource} and Eq. \ref{Bridgeshot} respectively. The source noise SNR is \begin{equation} \label{BridgesourceSNR} \textrm{SNR}_{\textrm{B,so}} = \frac{2\sin2\theta\sin2\theta_F}{\mathcal{B}\sqrt{\Delta f}\sqrt{1+\cos4\theta\cos4\theta_F}}, \end{equation} while the shot noise SNR is \begin{equation} \label{BridgeshotSNR} \textrm{SNR}_{\textrm{B,sh}} = \frac{\sqrt{2I_0}\sin2\theta\sin2\theta_F}{\sqrt{\Delta f}}. \end{equation}

\subsection{Comparison of the ideal SNR in the PCP, the optical bridge and the Sagnac interferometer}
\label{Comparison}

\begin{figure}[t]
\includegraphics[clip,width=8 cm] {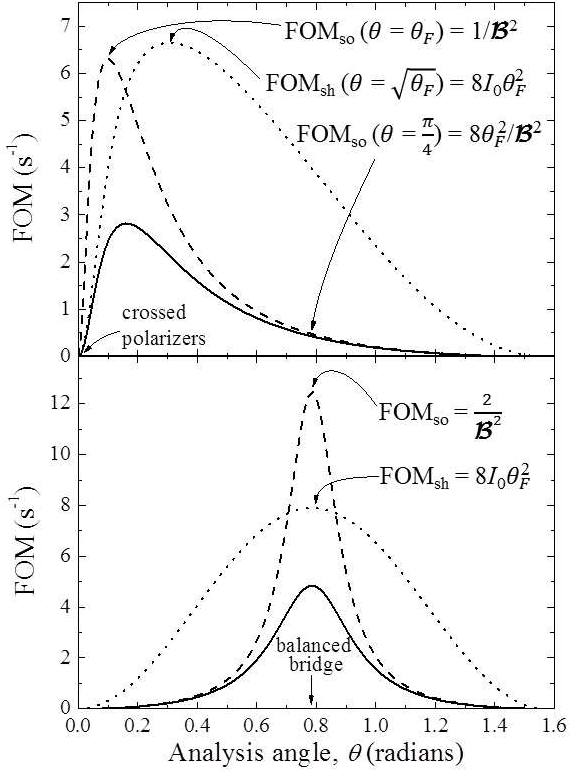}
\caption{Source (dashed lines) and shot (dotted lines) noise FOM calculated for the PCP/Sagnac case (top) and for the bridge (bottom) for $I_0 = 100$, $\mathcal{B} = 0.4$ and $\theta_F = 0.1$ rad using Eq. \ref{PCPsourceSNR}, Eq. \ref{PCPshotSNR}, Eq. \ref{SagnacshotSNR} ($\times$8), Eq. \ref{BridgesourceSNR} and Eq. \ref{BridgeshotSNR} with $\Delta f$ = 1 Hz. The total FOM (solid line) is calculated using Eq. \ref{totalFOM}. Maximum source and shot FOM occur close to the crossed polarizer (PCP) or dark fringe (Sagnac) condition when $\theta_F \ll 1$, but do not occur at the same analysis angle. Moreover the optimum analysis angle depends on $\theta_F$ and is therefore different for each sample. The optical bridge FOMs both show maxima at the balanced bridge condition, $\theta = \pi/4$. At this angle, the correlated source noise on the two bridge detectors results in a bridge to PCP source noise rejection ratio of 1/$4\theta_F^2$.}
\label{theory}
\end{figure}

In order to compare the three methods described above for given values of $\theta$, $\theta_F$, $I_0$ and $\mathcal{B}$, it is useful to define a figure-of-merit (FOM) for the source and for the shot noises, here denoted FOM$_\textrm{so}$ and FOM$_\textrm{sh}$ respectively. These FOM are defined as the squared SNR per unit bandwidth and are given by the product of $\Delta f$ with the appropriate squared SNR expression obtained from one of Eqns. \ref{PCPsourceSNR}, \ref{PCPshotSNR}, \ref{SagnacshotSNR}, \ref{BridgesourceSNR} or \ref{BridgeshotSNR}. The total FOM is thus expressed in units of $\textrm{Hz}$ (i.e. s$^{-1}$) and is given by \begin{equation} \label{totalFOM} \frac{1}{\textrm{FOM}}=\frac{1}{\textrm{FOM}_\textrm{so}}+\frac{1}{\textrm{FOM}_\textrm{sh}}. \end{equation} Fig. \ref{theory} shows the theoretical FOM curves versus analysis angle, $\theta$, for the source and shot noises in the PCP and the Sagnac (top panel) and the bridge (bottom panel) with $I_0 = 100$, $\theta_F = 0.1$ rad and $\mathcal{B} = 0.4$. These values are chosen so that the source and shot noise FOM are of similar magnitude thereby better revealing their $\theta$ dependence. In the case of the PCP/Sagnac shot noise FOM, Eqs. \ref{PCPshotSNR} and \ref{SagnacshotSNR} are maximized for $\theta = \sqrt{\theta_F}$ whereas Eq. \ref{PCPsourceSNR} for the source noise FOM is maximized when $\theta = \theta_F$. The total FOM peaks somewhere between these angles depending on the relative importance of the two components. In principal therefore an optimal alignment of the PCP or the Sagnac requires prior knowledge of the limiting noise source \textit{and} $\theta_F$. In addition, for nanoradian rotations a challenging  mechanical alignment to a precision in the $\theta_F$ to $\sqrt{\theta_F}$ range is required. As a consequence of this challenge, when using PCP it is usual to align to $\theta = \pi/4$ where the MO signal, Eq. \ref{PCPcontrast}, is maximized but where the FOM is lower. As shown in in Fig. \ref{theory} and in Table \ref{comparison}, in a shot noise limited measurement (for example at small $I_0$) this only reduces the FOM to half its maximum value, but for a source noise limited measurement a potentially significant reduction by a factor of $8\theta_F^2$ occurs.

\begin{table}
\footnotesize
    \begin{tabular}{| l | c | c | c | c | c | c |}
    \hline
    \textrm{FOM} & \specialcell{{\footnotesize PCP}\\ {\footnotesize $\theta = \sqrt{\theta_F}$}} & \specialcell{{\footnotesize Sagnac}\\ {\footnotesize $\theta = \sqrt{\theta_F}$}} & \specialcell{{\footnotesize PCP} \\ {\footnotesize $\theta = \theta_F$}} & \specialcell{{\footnotesize Sagnac} \\ {\footnotesize $\theta = \theta_F$}} & \specialcell{{\footnotesize PCP} \\ {\footnotesize $\theta = \frac{\pi}{4}$}} & \specialcell{{\footnotesize Bridge} \\ {\footnotesize $\theta = \frac{\pi}{4}$}} \\ \hline
    Shot & $8I_0\theta_F^2$ & $I_0\theta_F^2$ & $4I_0\theta_F^2$ & $I_0\theta_F^2/2$ & $4I_0\theta_F^2$ & $8I_0\theta_F^2$\\ \hline
    Source & $8\theta_F/\mathcal{B}^2$ & $8\theta_F/\mathcal{B}^2$ & $1/\mathcal{B}^2$ & $1/\mathcal{B}^2$ & $8\theta_F^2/\mathcal{B}^2$ & $2/\mathcal{B}^2$ \\ 
    \hline
    \end{tabular}
    \caption{A comparison of FOM for the three experimental geometries when $\theta_F \ll 1$ at particular values of $\theta$.}
  \label{comparison}
\end{table}

The bottom panel of Fig. \ref{theory} shows the FOM curves versus $\theta$ for the optical bridge. In contrast to the PCP/Sagnac case, here the maximum source and shot noise FOM occur at a sample independent $\theta = \pi/4$. Table \ref{comparison} shows a comparison of the source and shot noise FOM for the three configurations when $\theta_F \ll 1$ or $\theta = \theta_F$ where the source noise FOM is maximized for the PCP/Sagnac, at $\theta = \sqrt{\theta_F}$ where the shot noise FOM is maximized for the PCP/Sagnac, and at $\theta = \pi/4$ where the source and shot noise FOMs in the bridge are maximized. In this so-called balanced bridge condition, the optical source noise FOM is increased by a factor of $1/4\theta_F^2$ with respect to the PCP/Sagnac at $\theta=\pi/4$ (see also Fig. \ref{theory}). This permits a shot noise limited measurement even with large photon fluxes incident on the individual detectors, a significant advantage because the absolute value of the FOM increases with $I_0$ in this limit. Notice however that if the PCP is correctly aligned to $\theta = \sqrt{\theta_F}$, in the shot noise limit its FOM is identifcal to that of the bridge.

\subsection{Experimentally measured figures of merit}
\label{Experiment}

\begin{figure}
\includegraphics[clip,width=8 cm] {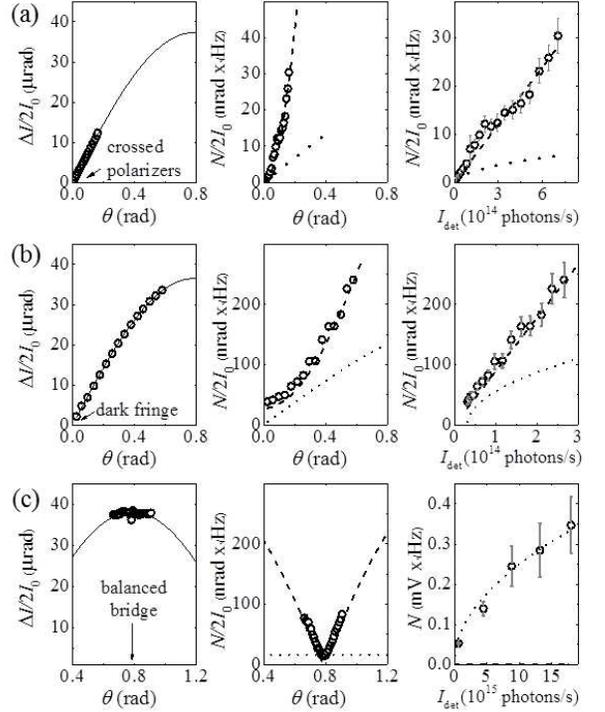}
\caption{Experimental results obtained using (a) the PCP, (b) the Sagnac, and (c) the bridge. The left panels show the measured MO signal (open circles) as a function of $\theta$, along with the fit and normalization made according to the corresponding equation for $\Delta I$ in each case, and with $\theta_F =$ 38 $\mu$rad and $I_0 \approx 2.3\times 10^{16}$ photons/s (PCP), $I_0 \approx 7.5\times 10^{15}$ photons/s (Sagnac) and $I_0 \approx 2.4\times 10^{16}$ photons/s (bridge). The middle panels show the noise measured as a function of $\theta$ on the MO signal (open circles) with the source (dashed line) and shot (dotted line) noise also shown for the experimental bandwidth $\Delta f =$ 53 Hz. The source noise parameter $\mathcal{B}$ is $10^{-7}$ (PCP), $0.68 \times 10^{-7}$ (Sagnac) and $0.3 \times 10^{-7}$ (bridge). The right panels show the $I_{\textrm{det}}$ dependence of the noise, revealing either the source (linear variation, dashed line) or shot (square root variation, dotted line) noise limited nature of the measurements.}
\label{experiment}
\end{figure}

\begin{figure*}
\includegraphics[clip,width=16 cm] {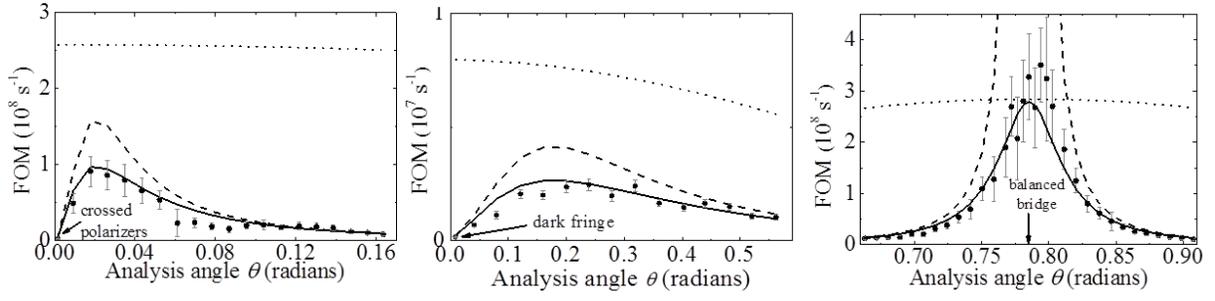}
\caption{The experimentally measured FOM for the PCP (left), Sagnac (middle) and bridge (right) (filled, black circles). With $I_0 \approx 5\times10^{15}$ photons/s in each case, the bridge provides the highest FOM since it is able to operate in the photon shot noise limit even at these (relatively) high detected intensities. The PCP is fully source noise limited after correction for the finite extinction ratio of the analyzer, while the Sagnac contains both a source and shot noise component. Although the FOM is lower, the shot noise component is more important in the Sagnac than in the PCP because the detected intensities are lower due to photon losses at the beam splitters.}
\label{FOMexperiment}
\end{figure*}

\begin{figure*}[t]
\includegraphics[clip,width=16 cm] {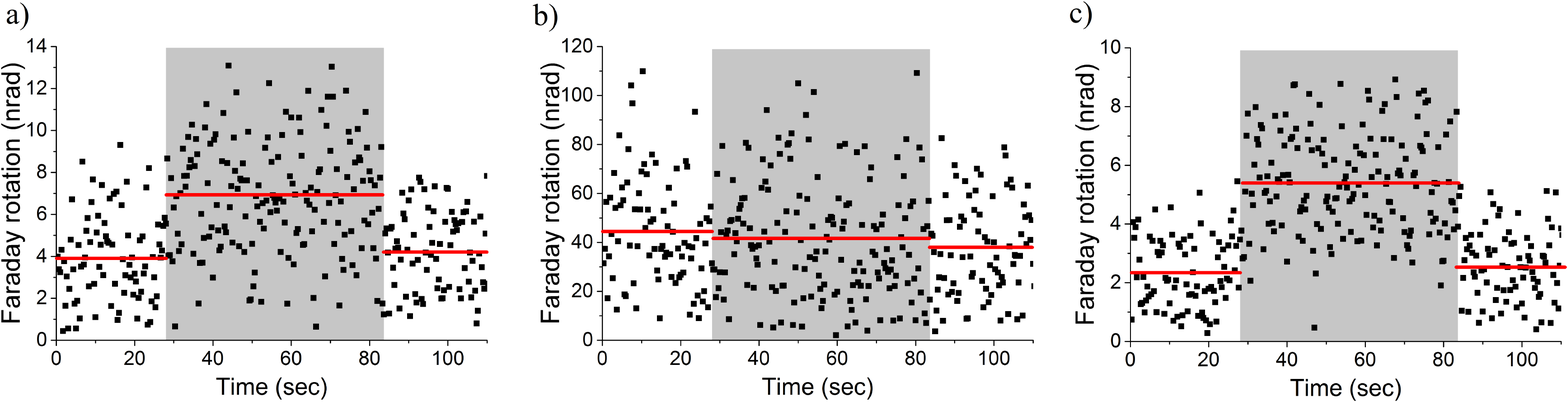}
\caption{Best case measurement of a 3 nrad polarization rotations for (a) PCP with $\theta = 0.03$, (b) the Sagnac interferometer with $\theta = 0.2$, and (c) the optical bridge with $\theta = \pi/4$. In each case, the vertical axis is $\theta_F$ determined from Eq. \ref{PCPcontrast}, Eq. \ref{Sagnaccontrast} or Eq. \ref{Bridgecontrast} respectively, using the measured MO signal, $\Delta I$, the known source intensity, $I_0$.}
\label{3nrad}
\end{figure*}

In order to demonstrate the capabilities of each of the three experimental configurations discussed here, the MO signal, the experimental noise and the resulting FOM are measured for a Faraday rotation angle, $\theta_F = 38$ $\mu$rad, established by applying a time varying magnetic field (of frequency 850 Hz) to a TGG of length 25 mm mounted in a solenoid, and then demodulating the resulting signal using a standard lock-in technique. A variable power (1-20 mW) 532 nm source is used and the experimental bandwidth is 53 Hz, corresponding to $2.7\times 10^{15} < I_0 < 5.3\times 10^{16}$ photons/s. The left panels of Fig. \ref{experiment} show the normalized  MO signal (open circles) measured using the PCP, Sagnac and bridge configurations respectively. The normalization factor is chosen in each case according Eq. \ref{PCPcontrast} (PCP), Eq. \ref{Sagnaccontrast} (Sagnac) and Eq. \ref{Bridgecontrast} (bridge). The functional equivalence of the PCP and Sagnac is apparent since they respectively follow the $\theta$ dependence of Eq. \ref{PCPcontrast} and Eq. \ref{Sagnaccontrast} (black lines in the left panels of Fig. \ref{experiment}(a) and Fig.\ref{experiment}(b) respectively). The MO signal on the bridge follows that predicted by Eq. \ref{Bridgecontrast}, represented as the black line in the left panel of Fig. \ref{experiment}(c).

In each of the experimental configurations the RMS noise on the MO signal is estimated by subtracting the mean MO signal from a statistically significant number of measurements, and then calculating the standard deviation of the resulting points. This is done as a function of the analysis angle as shown (open circles) in the middle panels of Fig. \ref{experiment} for the PCP, Sagnac and bridge configurations respectively. In these plots the normalization factors are the same as those used for the MO signal itself. In the (relatively high) intensity range used here, the only experiment that permits a photon shot noise limited measurement with a high FOM is the optical bridge because of the bridge to PCP source noise rejection discussed above (see Table \ref{comparison} and Fig. \ref{theory}). The measured RMS noise near the balanced bridge condition in the right panel of Fig. \ref{experiment}(c) is approximately equal to the shot noise calculated using Eq. \ref{Bridgeshot} (dotted line in the figure). Note that the estimation of the photon shot noise contains no adjustable parameters since the source intensity is known, $I_0 \approx 2.4\times 10^{16}$ photons/s. Away from the balanced condition, the optical source noise is no longer fully rejected (see the $\theta$ dependence of $N_{\textrm{B,so}}$ in Eq. \ref{Bridgesource} as plotted in the bottom panel of Fig. \ref{theory}). The right panels of Fig. \ref{experiment} shows the variation in the RMS noise on the MO signal with $I_{\textrm{det}}$. In the PCP and Sagnac a linear response characteristic of optical source noise is seen, while for $\theta = \pi/4$ in the bridge a clear $\sqrt{I_{\textrm{det}}}$ dependence again demonstrates the photon shot noise limited nature of this experiment. Notice that the noise in the bridge measurement is not normalized to $I_0$. Unlike the PCP and Sagnac cases where $I_{\textrm{det}}$ can be varied by fixing $I_0$ and varying the analysis angle, it can only be varied in the bridge by varying $I_0$. In this scenario if the $I_0$ normalization had been applied a $1/\sqrt{I_{\textrm{det}}}$ dependence would have resulted. The shot noise limited nature of the bridge measurement is clearer therefore without the $I_0$ normalization.

A photon shot noise limited measurement can be performed with the PCP or the Sagnac by reducing $I_0$ below some critical value since the optical source noise diminishes more rapidly with $I_0$ than the photon shot noise. In the limit $\theta_F \to 0$ near the crossed polarizers (or dark fringe) condition, $\theta \to 0$, the critical value at which the RMS source and shot noise become equal with PCP is found by equating Eq. \ref{PCPsource} with Eq. \ref{PCPshot} to obtain $I_{\textrm{crit}} = 1/\mathcal{B}^2$. In the Sagnac, because the RMS shot noise is a factor of $\sqrt{8}$ smaller for a given $I_0$ due to photon loss at the beam splitters, $I_{\textrm{crit}}$ is $\sqrt{8}$ times higher. Using the approximate values of $\mathcal{B}$ measured here ($10^{-7}$), a shot noise limited measurement could be expected with the PCP once the 532 nm light source power is reduced below $\approx$ 37 $\mu$W. This was confirmed experimentally (data not shown). In the case of the Sagnac, since light is lost at the beam splitters this limit is a factor of $\sqrt{8}$ higher ($\approx$ 0.1 mW). However, a reduction in light power is not usually desirable since (in the photon shot noise limit) it reduces the SNR.

The experimentally measured FOM for each configuration is shown in Fig. \ref{FOMexperiment} along with the source, shot and total FOMs. The right panel shows the photon shot noise (dotted line) limited measurement obtained with the balanced bridge clearly, as well as the renewed importance of the source noise (dashed line) away from this balanced condition where it is not properly compensated. As mentioned earlier the source (and therefore the total) noise is minimized at the balanced bridge condition (see Eq. \ref{Bridgesource}) so much so that an experimental estimation of the RMS noise becomes more difficult. This is manifested by the relatively large error bars near $\theta = \pi/4$ in the right panel of Fig. \ref{FOMexperiment}. The total FOM (solid line) is close to the measured FOM for all values of $\theta$ and contains only one fitting parameter, $\mathcal{B} = 0.68 \times 10^{-7}$. Although the $I_0$ is essentially identical for the bridge and the PCP, the bridge FOM is slightly higher than the PCP FOM (left panel of Fig. \ref{FOMexperiment}) because a shot noise limited measurement was not possible with the PCP at this light intensity. 

The functional equivalence of the PCP and the Sagnac is apparent from the similarities in the shapes of the experimental FOM curves shown in the left and middle panels of Fig. \ref{FOMexperiment} respectively. In both these cases the FOM is mainly limited by the optical source noise \textit{after} correction for the finite extinction ratio of the polarizers used in the PCP, or equivalently the finite fringe visibility in the Sagnac. A measured extinction ratio of $4 \times 10^{-4}$ in the PCP and a fringe visibility of 0.92 (corresponding to an effective extinction ratio of $4 \times 10^{-2}$) in the Sagnac are accounted for by adding an offset intensity to Eq. \ref{Malus} and to Eq. \ref{SagnacMalus} respectively. This then yields a corrected source noise FOM represented by the dashed lines in the left and middle panels of Fig. \ref{FOMexperiment}. Note that other undesired offsets in the measured intensity may arise due to physical rotation of the Sagnac loop, even that due to the Earth’s rotational movement. However the magnitude of this offset scales with the effective loop area and can be removed in zero-area fibered Sagnac interferometers \cite{fried2014}.The Sagnac FOM is $\approx$ 30 times lower than the PCP FOM, in part because of the factor of 8 reduction in $\textrm{FOM}_\textrm{sh}$ for the Sagnac relative to the PCP, and in part because $I_0$ is a factor of 3 lower in the experimental realization of the Sagnac i.e. a total factor of $8 \times 3 = 24$. The remaining difference arises because the effective extinction ratio of the Sagnac is significantly poorer than that of the PCP due to back reflection of non-interfering photons off the numerous optical components in the interferometer. The difference in extinction ratios is also responsible for the fact that the maximum source noise FOM occurs at an analysis angle ($\theta \neq \theta_F$) which is different in each case.  

The superior experimental FOM of the bridge is demonstrated by measuring a 3 nrad polarization obtained by applying an average axial field of 0.9 nT to the TGG. Fig. \ref{3nrad}(c) clearly demonstrates a SNR greater than 1 when using the bridge with a bandwidth of $\Delta f = 1.5$ Hz (the external magnetic field is applied for a time corresponding to the gray zone). Here the vertical axis corresponds to $\theta_F$ obtained by using the appropriate MO signal expression for each of the experimental configurations at the analysis angle yielding the maximum FOM (i.e. Eq. \ref{PCPcontrast} for the PCP with $\theta = 0.03$, Eq. \ref{Sagnaccontrast} for the Sagnac with $\theta = 0.2$, and Eq. \ref{Bridgecontrast} for the bridge with $\theta = \pi/4$). Under identical conditions (i.e. bandwidth, source intensity) the 3 nrad rotation is just visible with the PCP but well within the noise in the Sagnac measurement. Although the differences in the FOM between configurations are small, in the absence of any other constraints, the bridge is therefore the preferred configuration because it allows for a photon shot noise limited measurement at high detected intensities where the SNR is large.

\section{Discrimination of time and parity symmetries with Sagnac interferometers}
\label{SagnacGeometries}

\begin{figure*}[t]
\includegraphics[clip,width=16 cm] {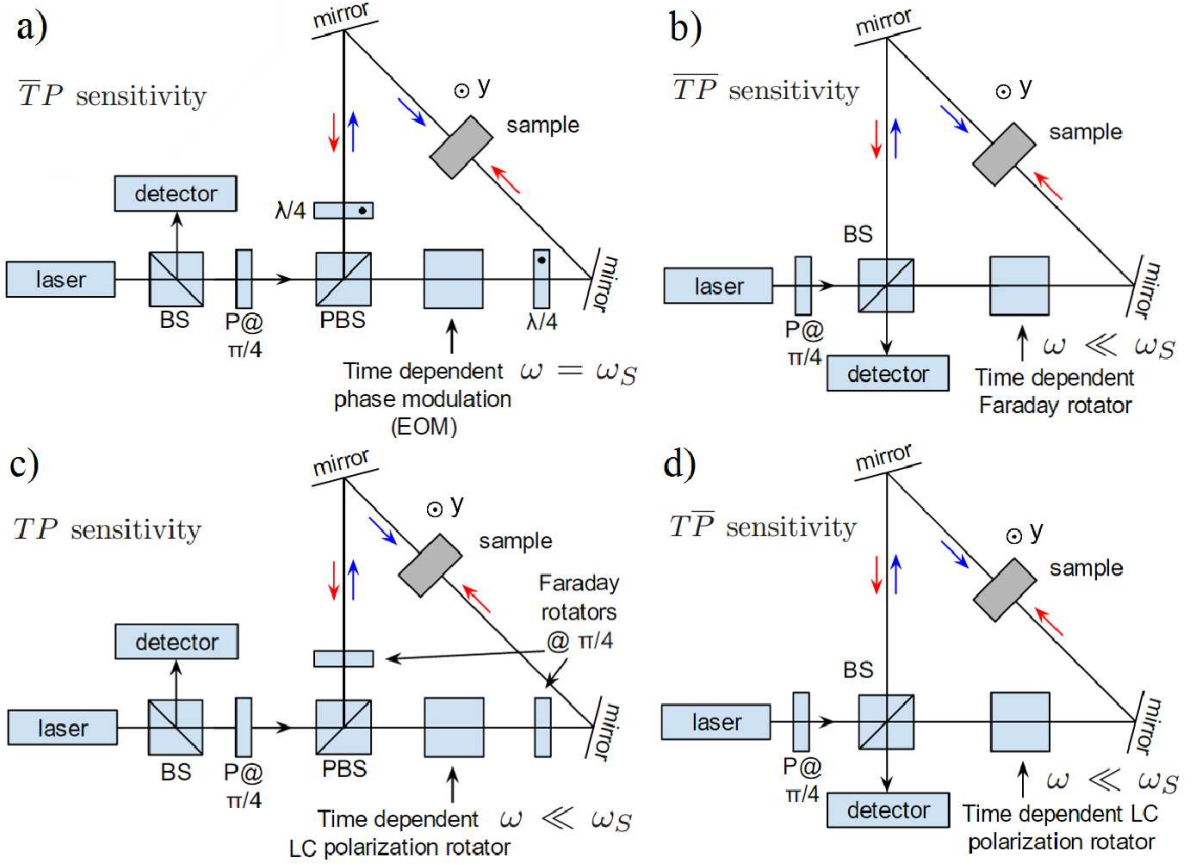}
\caption{Four Sagnac interferometers, each sensitive only to polarization rotations with specific time and spatial symmetries. (a) $\overline{T}P$ sensitivity only as implemented and used elsewhere \cite{xia2006}, (b) $TP$ sensitivity only, (c) $\overline{TP}$ sensitivity only, and (d) $T\overline{P}$ sensitivity only. In each case the acronyms given for the optical components are the same as those given in Fig. \ref{configs}. In the case of the linear polarizers, ``P'', the angle following the label corresponds to the angle the optical axis makes with the x-axis.}
\label{Sagnacs}
\end{figure*}

Although the Sagnac FOM suffers from the loss of potentially useful photons at the two beam splitters, it has been used to distinguish polarization rotations due to the non-reciprocal ($\overline{T}$) Faraday effect from those due to reciprocal ($T$) linear birefringence \cite{spielman1990}. While this is in principle possible with PCP or with the bridge, it would require at least two independent measurements (for example, in which the sample is measured once, rotated by 180$^{\circ}$ around an axis perpendicular to the light propagation direction and then re-measured). Two independent measurements are often difficult to achieve under identical experimental conditions, or in a spatially inhomogeneous sample. The common path nature of the Sagnac ensures that the same part of sample will be measured simultaneously by both the CW and CCW beams.  

The well known ability of Sagnac interferometers to distinguish rotations arising from reciprocal and non-reciprocal phenomena can be understood by noting that (to within a phase conjugation) the reversal of the light wavevector describing the CW and CCW beams, $k_z \to -k_z$, corresponds to an effective time reversal of the electromagnetic wave: $\mathcal{E}(z,t) \propto e^{i(\omega t - k_zz)}$ so that $\mathcal{E}(z,-t) \propto e^{i(-\omega t + k_zz)} = \mathcal{E}^*(-z,t)$.

While not yet widely discussed, Sagnac interferometers are also sensitive to mirror symmetry across the $xy$ plane of the sample. This can be understood by realizing that the order of the optical elements in a Sagnac loop affects its function. If the sample is modeled as two plates separated by the $xy$ plane then, in an $xy$ mirror symmetric case (and only in this case) the overall optical response (as described by the product of the Jones' matrices of the two plates) to CW and CCW beams will be identical. Thus the sample's mirror symmetry (or lack thereof) in the $xy$ plane can be detected with a Sagnac. From the point of view of the constraints on the symmetry of the Jones' matrix of the sample, parity and $xy$ mirror symmetry are equivalent \cite{armitage2014}. In the following ``parity'' ($P$) will, strictly speaking, refer only to the inversion operation along the $z$ axis of the sample.

To date, the literature contains only variants of a Sagnac interferometer that is designed to distinguish between non-reciprocal polarization rotations induced by magnetism (the Faraday effect, $\overline{T}P$ symmetry) from reciprocal, even parity rotations ($TP$) induced by linear birefringence \cite{xia2006}, including the ability to (vector)  discriminate between polar, longitudinal and transverse magneto-optical Kerr effects \cite{dodge1996}. This type of geometry is shown in Fig. \ref{Sagnacs}(a). In fact as will be shown here, this geometry is also insensitive to $\overline{TP}$ and $T\overline{P}$ rotations, for example those arising from ferroelectricity and from chirality respectively. Moreover, it will be seen that Sagnac interferometers can also be designed to be sensitive \textit{only} to polarization rotations arising from phenomena with these other time and parity symmetries. In order to understand the functionality of the example Sagnac interferometers proposed here, the generalized Jones matrix symmetries and mapping rules for each of these time and spatial symmetries will be recalled \cite{armitage2014}. In the forward propagating direction which may be arbitrarily assigned to the CW beam, the Jones' matrices for each of these four combinations are, for a lossless material, given by:
\begin{align} \label{eq:forw}
\textbf{M}_{TP} &= \begin{bmatrix} A&B\\ B&D\\ \end{bmatrix} & \textbf{M}_{\overline{TP}} &= \begin{bmatrix} A&B\\ B&D\\ \end{bmatrix} \notag \\
\textbf{M}_{T\overline{P}} &= \begin{bmatrix} A&-B\\ B&D\\ \end{bmatrix} & \textbf{M}_{\overline{T}P} &= \begin{bmatrix} A&-B\\ B&D\\ \end{bmatrix}.
\end{align} Notice that $\textbf{M}_{\overline{T}P}$ has the form initially given for the Faraday effect itself, Eq. \ref{FaradayJones}. In the reverse or CCW direction, corresponding to backside transmission of light through the sample, these map to \cite{armitage2014}:
\begin{align} \label{eq:back}
\hat{\textbf{M}}_{TP} &= \begin{bmatrix} A&-B\\ -B&D\\ \end{bmatrix} & \hat{\textbf{M}}_{\overline{TP}} &= \begin{bmatrix} A&B\\ B&D\\ \end{bmatrix} \notag \\
\hat{\textbf{M}}_{T\overline{P}} &= \begin{bmatrix} A&-B\\ B&D\\ \end{bmatrix} & \hat{\textbf{M}}_{\overline{T}P} &= \begin{bmatrix} A&B\\ -B&D\\ \end{bmatrix}.
\end{align} Here the caret symbol over the matrix operator is used to denote backside transmission \cite{armitage2014}. These forms will be used for the calculations of the responses of each of the Sagnac loops shown in Fig. \ref{Sagnacs}. Note again that the Jones' matrices are defined in the reference frame of the light so that the z-axis of the matrices in Eq. \ref{eq:back} is reversed relative to that of the matrices in Eq. \ref{eq:forw}. It is for this reason that the $\overline{T}P$ matrices describing (amongst others) the Faraday effect are identical for forwards and backwards transmission while the there is a change of sign of the off-diagonal components of the $T\overline{P}$ matrices (i.e. for chirality). Written in the reference frame of the laboratory the situation would have been reversed. The same arguments hold for the $TP$ and $\overline{TP}$ cases.

Let us first consider the configuration given in Fig. \ref{Sagnacs}(a) which has been used elsewhere \cite{kapitulnik1994, xia2006}. In these works, the Sagnac loop is similar to that used here (see Fig. \ref{configs}(b)) but with the static Faraday rotator removed, and an electro-optic phase modulator (EOM) introduced with its fast axis aligned along the x-direction. The linearly birefringent ($TP$) EOM is a critical element that gives the Sagnac its ability to distinguish polarization rotations of different symmetries by inducing a time dependent phase delay between the two components of the light according to $\textbf{PR}(\phi(t),0,0)$ where $\phi(t) = \phi_{0}\sin\omega t$. The frequency of this phase modulation is chosen so that $\omega = \pi/\tau_S = \omega_S$, where $\tau_S$ is the propagation time for light in the Sagnac loop (in a fibered loop of the order of several hundred metres long, $\omega_S$ falls in the MHz range). In this case, for a CW wavefront encountering a phase lag of $\phi_{0}\sin\omega_S t^\prime$ at the position of the EOM in the loop at a time $t = t^\prime$, the CCW wavefront at the EOM encounters a phase lag of $\phi_{0}\sin(\omega_S (t^\prime+\tau_s))=-\phi_{0}\sin\omega_S t^\prime$. The intensity on the detector with a sample that induces a Faraday rotation $\theta_F$ is then given by Eq. \ref{SagnacMalus} with the constant analysis angle, $\theta$, replaced by $\phi_{0}\sin\omega_S t$. By choosing $\omega = \omega_S$ the EOM phase retardation becomes effectively non-reciprocal despite the $T$ symmetry of the EOM itself. The harmonics of this time dependent signal can then be measured and the DC, $\omega$ (first harmonic) and $2\omega$ (second harmonic) components are given in Table \ref{harmonics}.

\begin{table*}
		\footnotesize
    \begin{tabular}{| c | c | c | c |}
    \hline
     & DC & $\omega$ & $2\omega$ \\ \hline
    Fig. \ref{Sagnacs}(a) &  &  &  \\
    $TP$ & $I_0 (A+D)^2(1+J_0(2\phi_0))/16$ & 0 & $I_0(A+D)^2J_2(2\phi_0)/8$\\
    $\overline{TP}$ & $I_0 (A+D)^2(1+J_0(2\phi_0))/16$ & 0 & $I_0(A+D)^2J_2(2\phi_0)/8$\\
    $\overline{T}P$ & $I_0 \left((A+D)^2+4B^2+((A+D)^2-4B^2)J_0(2\phi_0)\right)/16$ & $-I_0B(A+D)J_1(2\phi_0)/2$ & $I_0\left((A+D)^2 - 4B^2\right)J_2(2\phi_0)/8$\\
    $T\overline{P}$ & $I_0 (4B^2+(A+D)^2)(1+J_0(2\phi_0))/16$ & 0 & $I_0\left((A+D)^2 + 4B^2\right)J_2(2\phi_0)/8$\\ \hline
    Fig. \ref{Sagnacs}(b) &  &  &  \\ 
    $TP$ & $I_0 \left(4B^2 +(A-D)^2\right)(1-J_0(2\phi_0))/16$ & 0 & $-I_0  (4B^2+(A-D)^2) J_2(2\phi_0)/8$\\
    $\overline{TP}$ & $I_0 \left((A-D)^2 + 4B^2 - ((A-D)^2-4B^2)J_0(2\phi_0)\right)/16$ & $I_0 B(D-A) J_1(2\phi_0)/2$ & $-I_0 \left(-4B^2+(A-D)^2\right) J_2(2\phi_0)/8$ \\
    $\overline{T}P$ & $I_0 (A-D)^2(1-J_0(2\phi_0))/16$ & 0 & $-I_0(A-D)^2 J_2(2\phi_0)/8$\\
    $T\overline{P}$ & $I_0\left(8B^2 - (D-A)(A+4B-D)(1+J_0(2\phi_0))\right)/16$ & 0 & $I_0(D-A)(A+4B-D) J_2(2\phi_0)/8$\\ \hline
    Fig. \ref{Sagnacs}(c) &  &  &  \\ 
    $TP$ & $I_0 \left((A-D)^2 + 4B^2 - ((A-D)^2-4B^2)J_0(2\phi_0)\right)/16$ & $I_0 B(D-A) J_1(2\phi_0)/2$ & $-I_0\left((A-D)^2-4B^2\right) J_2(2\phi_0)/8$\\
    $\overline{TP}$ & $I_0 (A-D)^2(1-J_0(2\phi_0))/16$  & 0 & $-I_0 (A-D)^2 J_2(2\phi_0)/8$\\
    $\overline{T}P$ & $I_0 (-A+2B +D)^2(1-J_0(2\phi_0))/16$ & 0 & $-I_0(-A+2B+D)^2 J_2(2\phi_0)/8$\\
    $T\overline{P}$ & $I_0 (A-D)^2(1-J_0(2\phi_0))/16$ & 0 & $-I_0(A-D)^2 J_2(2\phi_0)/8$\\ \hline
    Fig. \ref{Sagnacs}(d) &  &  &  \\  
    $TP$ & $I_0 (A+D)^2(1-J_0(2\phi_0))/16$ & 0 & $-I_0(A+D)^2 J_2(2\phi_0)/8$\\
    $\overline{TP}$ & $I_0\left(8B^2 + (A+D)(A+4B+D)(1-J_0(2\phi_0))\right)/16$ & 0 & $-I_0(A+D)(A+4B+D) J_2(2\phi_0)/8$\\
    $\overline{T}P$ & $I_0 (4B^2+(A+D)^2)(1-J_0(2\phi_0))/16$ & 0 & $-I_0 \left(4B^2+(A+D)^2\right)J_2(2\phi_0)/8$\\
    $T\overline{P}$ & $I_0 \left((A+D)^2+4B^2-((A+D)^2-4B^2)J_0(2\phi_0)\right)/16$ & $I_0 B (A+D) J_1(2\phi_0)/2$ & $-I_0\left((A+D)^2-4B^2\right) J_2(2\phi_0)/8$ \\ \hline
    \end{tabular}
    \caption{Harmonic components of the time dependent intensity measured on the detector in each of the four Sagnac interferometers shown in Fig. \ref{Sagnacs}.}
  \label{harmonics}
\end{table*}

At first sight Table \ref{harmonics} looks quite imposing, but consider the case of a pure Faraday rotation, $A = D = \cos\theta_F$ and $B = \sin\theta_F$ applied to the $\overline{T}P$ response of the Sagnac interferometer sensitive to this symmetry (i.e. that of Fig. \ref{Sagnacs}(a)). The $\omega$ component of the measured intensity simplifies to $-I_0\sin 2\theta_F J_1(2\phi_0)/2$ while the $2\omega$ component becomes $I_0 \cos 2\theta_F J_2(2\phi_0)/2$. Here $J_1$ and $J_2$ are Bessel functions. The magnitude of their ratio is then proportional to $\tan 2\theta_F$ \cite{kapitulnik1994} and independent of $I_0$ so that $\theta_F$ can be directly determined by measuring this ratio. This technique is more generally known as the Phase Generated Carrier (PGC) method \cite{dandridge1982}. Notice that the $\omega$ component of all rotations arising from phenomena with other symmetries is $0$ so that the ratio of first to second harmonics is only non-zero for $\overline{T}P$ rotations. It should also be noted that the DC components are non-zero \textit{in all cases}, so that the static version of the Sagnac loop in Fig. \ref{Sagnacs}(a) does not distinguish between the time and parity symmetries of the polarization rotation. Thus the loop implemented here (see Fig. \ref{configs}(b)) is sensitive to all types of rotation, although this does not affect its FOM.

With the Jones' matrix formalism it is possible to test other Sagnac configurations exploiting the PGC method in order to search for cases which are sensitive only to $\overline{TP}$, $T\overline{P}$ or $TP$ rotations. The response of the Sagnac loops in Fig. \ref{Sagnacs}(b), (c) and (d) are also given in Table \ref{harmonics} and by observing the $\omega$ components whose intensity is zero, it can be seen that sensitivity to \textit{only} $\overline{TP}$, $TP$ or $T\overline{P}$  rotations can be achieved respectively. Thus if one wished, for example, to test for the presence of magneto-electricity while avoiding responses due to standard magnetic phenomena, the Sagnac loop in Fig. \ref{Sagnacs}(b) could be used. 

\begin{figure}[t]
\includegraphics[clip,width=8 cm] {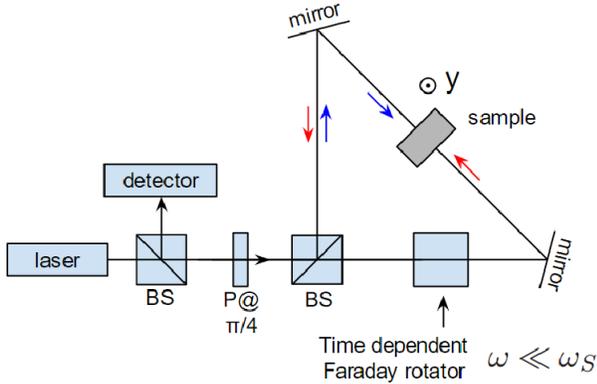}
\caption{An alternative Sagnac interferometer whose response is identical to that of the interferometer shown in Fig. \ref{Sagnacs}(a).}
\label{alternative}
\end{figure}

Like the $\overline{T}P$ case, the modulation component is crucial in the Sagnacs shown in Figs. \ref{Sagnacs}(b), (c) and (d). In the $\overline{TP}$ loop (Fig. \ref{Sagnacs}(b)) this component is a Faraday rotator, modulated at a frequency $\omega \ll \omega_S$ i.e. in the static limit. In practice a time modulated Faraday rotation can be achieved with a sufficiently long TGG placed in a solenoid. Thus for the CW beam this component induces a rotation given by $\textbf{F}(\phi_0\sin\omega t)$ while for the CCW beam the rotation is $\textbf{F}(-\phi_0\sin\omega t$) \cite{commentRSI1}. In the Sagnac loops in Fig. \ref{Sagnacs}(c) and (d), a time dependent liquid crystal (LC) modulator is used where $\omega \ll \omega_S$ i.e. again in the static limit. The LC rotator is a chiral ($T\overline{P}$) object, so that the CW and CCW beams see rotations given by $\textbf{F}(\phi_0\sin\omega t)$ \cite{commentRSI2} according to Eq. \ref{eq:forw} and Eq. \ref{eq:back} with $A = D = \phi_0\sin\omega t$ and $B = \phi_0\cos\omega t$. In practice, commercially available LC rotators cannot (for the moment) provide a sinusoidally modulated polarization rotation. Rather they are able to provide a low frequency square wave modulation between two fixed rotation values, which is still sufficient to obtain the functionality of the Sagnac loops in Fig. \ref{Sagnacs}(c) and Fig. \ref{Sagnacs}(d). It should also be noted that the example interferometers given here are not the only possible configurations that yield sensitivity to rotations with a given time and parity symmetry. As an example, the Sagnac interferometer in Fig. \ref{alternative} yields an identical response to that given in Fig. \ref{Sagnacs}(a). This is an interesting case because the requirement that $\omega = \omega_S$ in the loop shown in Fig. \ref{Sagnacs}(a) requires a loop length of at least several hundred metres since commercial EOMs can only be modulated up to the MHz range. Such a long loop is really only practicable with optical fibers. On the other hand the loop in Fig. \ref{alternative} can be implemented as a free space Sagnac since the requirement $\omega \ll \omega_S$ is easily achieved.

\section{Conclusion}
Three optical configurations permitting the measurement of polarization rotations have been studied. While in principal all three -- PCP, Sagnac interferometers and optical bridges -- can be used with equivalent FOM, a common mode rejection of optical source noise in the bridge allows it to perform measurements in the photon shot noise limit even at high detected intensities. The common mode rejection of the source noise is particularly efficient when nanoradian polarization rotations are to be measured. In terms of the FOM, the static Sagnac interferometer implemented and studied here is functionally equivalent to the PCP. In this static configuration, the Sagnac is incapable of distinguishing between rotations arising from phenomena with different time and parity symmetries. Replacement of the static Faraday rotator with an EOM operating at the loop frequency, $\omega_s$, yields a Sagnac which is sensitivity only to non-reciprocal rotations with $\overline{T}P$ symmetry (e.g. those due to the Faraday effect). While this configuration has been known for some time, it is shown here that Sagnac interferometers can also be configured to be sensitive $\textit{only}$ to rotations with other combinations of time and parity symmetries. Examples are given for each case, including configurations sensitive to purely reciprocal phenomena.

\appendix
\section{Generalized Jones' matrices phase retarders, polarizers, mirrors and non-polarizing beam splitters}
\label{Jones}

\subsection{Phase retarders}

The general form of the Jones' matrix for a phase retarding material is: \begin{align} \label{pr} &\textbf{PR}(\phi_x,\phi_y,q) = \notag \\ &\left[\begin{array}{cc} \textrm{e}^{i\phi_x}\cos^2q + {e}^{i\phi_y}\sin^2q & \left(\textrm{e}^{i\phi_x} - \textrm{e}^{i\phi_y}\right)\cos q \sin q \\ \left(\textrm{e}^{i\phi_x} - \textrm{e}^{i\phi_y}\right)\cos q \sin q & \textrm{e}^{i\phi_x}\sin^2q + {e}^{i\phi_y}\cos^2q \end{array} \right].\end{align} Here $q$ is the angle the fast axis of the retarder makes with the x-axis, $\phi_x$ is the phase delay introduced in the polarization component parallel to the x-axis, while $\phi_y$ is the equivalent phase delay for the y component of the polarization.

\subsection{Linear polarizers}

The general Jones' matrix expression for a linear polarizer is given by \begin{equation} \label{pol} \textbf{P}(\theta_P) = \left[\begin{array}{cc} \cos^2\theta_P & \sin\theta_P\cos\theta_P \\ \sin\theta_P\cos\theta_P & \sin^2\theta_P \end{array} \right].\end{equation} Here $\theta_P$ is the angle the polarizer axis makes with the x-axis.

\subsection{Mirrors}

In the calculations presented in this article, the Jones' matrix for a perfect metallic mirror at close to normal incidence is used: \begin{equation} \label{mirror} \textbf{M} = \left[\begin{array}{cc} 1 & 0 \\ 0 & -1 \end{array} \right].\end{equation} This expression shows that the x-component (the p polarized component in the coordinate system used here) is reflected in phase with the incident beam, while the y-component (the s polarized component in the coordinate system used here) is reflected in anti-phase with the incident beam. In all of the Sagnac loops presented in this article there are two mirrors present which yield $\textbf{M}^2 = \textbf{1}$ where $\textbf{1}$ is the $2\times2$ identity matrix.

\subsection{Non polarizing beam splitter}

The 50/50 BS used here is a commercially available model marked with an input face whose optical properties were characterized experimentally in order to determine the equivalent Jones' matrices. When light enters from the input face and is transmitted through the cube, the polarization is unaffected although the intensity is halved: \begin{equation} \label{bst} \textbf{BST} = \frac{1}{\sqrt{2}}\left[\begin{array}{cc} 1 & 0 \\ 0 & 1 \end{array} \right]. \end{equation} Light entering from the input face that is reflected is transformed like the mirror, Eq. \ref{mirror}, with a halved intensity: \begin{equation} \label{bsr} \textbf{BSR} = \frac{1}{\sqrt{2}}\left[\begin{array}{cc} 1 & 0 \\ 0 & -1 \end{array} \right]. \end{equation} In the Sagnac loop light may re-enter the 50/50 BS at the reflected output face in order to be transmitted towards the detector. This is the so-called anti-tranmission, and is given by $\textbf{ABST} = \textbf{BST}$. Similarly, light can re-enter the cube via the transmitted output face in which case it is reflected towards the detector according to $\textbf{ABSR} = -\textbf{BSR}$. The important point is that this so-called anti-reflection occurs in anti-phase.

\begin{acknowledgements}
The authors would like to thank T. Talvard and R. Crisan for their help in setting up parts of the experiment. The authors thank the Agence Nationale de la Recherche (DiracForMag, ANR-14-CE32-0003) for financial support.
\end{acknowledgements}



\begin{thebibliography}{31}
\bibitem[{com({\natexlab{a}})}]{commentRSI1}
\bibinfo{note}{If $\omega = \omega_S$ had been chosen the CCW beam would see a
  rotation given by $\textbf{F}(-\phi_0\sin(\omega t + \pi)) =
  \textbf{F}(\phi_0\sin\omega t)$, the same as the CW beam. In effect the time
  dependent Faraday rotator is made to behave as a chiral object.}
\end{thebibliography}

\end{document}